# Designing Sensing as a Service (S$^2$aaS) Ecosystem for Internet of Things

Charith Perera, *Cardiff University, United Kingdom (charith.perera@ieee.org)*
Mahmoud Barhamgi, *Claude Bernard University Lyon-1, France (mahmoud.barhamgi@univ-lyon1.fr)*
Suparna De, *University of Surrey, United Kingdom (s.de@surrey.ac.uk)*
Tim Baarslag, *Centrum Wiskunde & Informatica (CWI), Netherlands (T.Baarslag@cwi.nl)*
Massimo Vecchio, *FBK CREATE-NET, Italy (mvecchio@fbk.eu)*
Kim-Kwang Raymond Choo, *UTSA College of Business, United States (Raymond.Choo@utsa.edu)*

**ABSTRACT**

The Internet of Things (IoT) envisions the creation of an environment where everyday objects (e.g. microwaves, fridges, cars, coffee machines, etc.) are connected to the internet and make users' lives more productive, efficient, and convenient. During this process, everyday objects capture a vast amount of data that can be used to understand individuals and their behaviours. In the current IoT ecosystems, such data is collected and used only by the respective IoT solutions. There is no formal way to share data with external entities. We believe this is very efficient and unfair for users. We believe that users, as data owners, should be able to control, manage, and share data about them in any way that they choose and make or gain value out of them. To achieve this, we proposed the Sensing as a Service (S$^2$aaS) model. In this paper, we discuss the Sensing as a Service ecosystem in terms of its architecture, components and related user interaction designs. This paper aims to highlight the weaknesses of the current IoT ecosystem and to explain how S$^2$aaS would eliminate those weaknesses. We also discuss how an everyday user may engage with the S$^2$aaS ecosystem and design challenges.

**INTRODUCTION**

From a business perspective, IoT can be segmented into many different sectors such as smart home, smart city, smart wearables, smart manufacturing, and so on. These segments are not mutually exclusive. Smart home and smart wearable segments have received more attention, emphasis, and investment from the industry due to the potential value of data they collect.

Over the past few years, a large number of IoT solutions have come to the IoT marketplace [1], [2]. Typically, each solution is designed to perform a single or minimal number of tasks (primary usage). For example, a smart sprinkler may only be activated if the soil moisture falls below a certain level in a garden. Further, smart plugs allow users to control electronic appliances (including legacy appliances) remotely or create automated schedules. Such automation not only brings convenience to users but also reduces resource wastage (e.g. through efficient planning and predictions). Some other IoT solutions such as *Fitbit* and *Beddit* [1], [2] collects, and analyse information about individuals and present somewhat useful and summarised information back to the owner.

**WEAKNESSES IN THE CURRENT IOT**

The data collected by each IoT solution is only used by themselves and stored in access-controlled silos. After the primary usage, data is either thrown away or locked down in independent data silos. Each IoT solution aims to bring some value to our lives. For example, Fitbit *"motivates you to reach your health and fitness goals by tracking your activity, exercise, sleep, weight and more"*[1]. We acknowledge that such IoT solutions are useful. However, what if we want to combine and correlate our *Fitbit* data with *Beddit* data. Fitbit and *Beddit* are two different IoT solutions developed by two different companies. In today's IoT ecosystem, we do not have a formal way to combine data from different IoT solutions and perform analysis and reasoning over combined data sets. There is a significant amount of knowledge hidden in these independent IoT product silos that can be used to improve our lives (including behaviours, habits, and life patterns) and reduce wastage through efficient resource consumption.

---
[1] https://www.fitbit.com

On the other hand, there are also social and economic reasons to claim that the current IoT ecosystem is unfair and inefficient. We purchase different IoT devices. Therefore, we should own the data captured by these devices. We should be able to do whatever we like with the data, including sharing and trading them with entities we prefer, under our own terms. It is no secret that personal data has significant economic values [3]. In traditional Internet domain, as well as in mobile and social media domains, consumer data is being considered as a gold mine. *So why shouldn't we get to control our data and monetise it?* However, it is important to note that most of us (i.e. non-technical data owners) are not able to analyse our own data, due to lack of expertise. Therefore, most of the time, it will be required to share the data with external parties in order to get the data analysed, processed and useful knowledge derived.

**ADDRESSING WEAKNESSES IN THE IOT ECOSYSTEM**

Let us identify the requirements of an ideal solution that addresses the weakness mentioned above in the existing IoT ecosystem.

- **Data Trading** [4]**:** Data owners should be able to trade/share their data (captured by the IoT solutions they own) with the entities they prefer.
- **Market Place:** For data trading, there should be a meeting place where data owners and data consumers can meet with minimum or no cost.
- **Control:** Data sharing should happen under the conditions imposed by respective data owners. In the current mobile app ecosystem, users have no other option than to provide all the permissions requested by a given app. In an ideal data market place [5], data owners should be able to negotiate in details which data items to be traded and under which conditions [6].
- **Rewards:** Personal data has value [7]. Therefore, data owners should be able to receive rewards by sharing their data with external entities. Rewards [4] may have several shapes such as cash, points, bitcoins, vouchers, discounts, free goodies, and, more importantly, useful advices.
- **Ease of Use:** Tools and techniques need to be put in place in a way that even an average, non-technical, user can engage with the ecosystem. For example, the data owner needs to be properly informed regarding data trading activities, risks, and consequences involved.
- **Security and Privacy:** Data need to be secure at all times (during acquisition, storage, and analysis). Further, the individual data owner's privacy needs to assured.

**SENSING AS A SERVICE**

Sensing as a Service ($S^2$aaS) aims at addressing all weakness mentioned above, by demonstrating the above characteristics. It is a vision and a business model promoting data exchange (i.e. trading) between data owners and data consumers. Imagine a world where data owners (who own IoT solutions) are rewarded (e.g. money, loyalty points, gifts, vouchers, bitcoin, actionable advice, etc.) for sharing (i.e. trading) their data (collected by IoT products). From the other end, companies (i.e. data consumers) get to better understand their customers (i.e. data owners). As a result, companies will be able to optimize their business operations, by saving costs and to create new products and services better fitting individual customer needs. Data consumers may recover their data acquisition costs through business process optimization and increased customer (i.e. data owners) satisfaction. Data consumers can be governments or not-for-profits as well [8]. S$^2$aaS model is discussed in detail in [9]. It is important to note that the sensing as a service model is not limited to personal data sharing. Instead, it applies to smart cities as well, where data is not owned by individuals, but also by the organization, as well as the public (i.e., government).

However, the most dramatic change is expected to happen in the personal data domain. Therefore, in this paper, we primarily focus on personal data sharing. Let us explain the S$^2$aaS using the smart home scenario. Next, we will also discuss a smart city scenario.

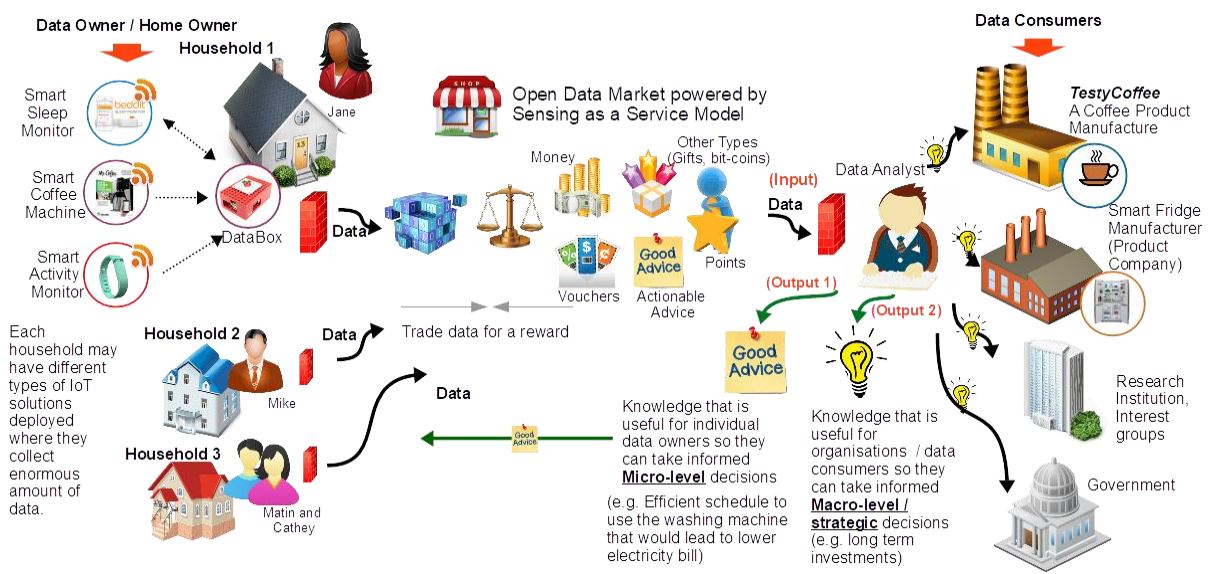

*Figure 1: Data Market for Sensing as a Service in Smart Home Domain*

*TastyCoffee* is a coffee products manufacture keen to know how people like Jane consume coffee (e.g. patterns, amounts, etc.). Jane is a restaurant manager living alone in her own house. She has three different IoT products in her house, as illustrated in Figure 1. *TastyCoffee* wants to know whether there are any relationships between activity patterns, quality of sleep, and coffee consumption (different variations). For example, *TastyCoffee* would like to discover any consumer patterns (e.g., whether people tend to drink less coffee on a day with fewer workouts). Currently, the only way that they could discover this kind of information is through user surveys and focus group studies. However, such methods are time-consuming, inaccurate and expensive to carry out. However, if *TastyCoffee* could access Jane's silo (and thousands of other similar users) consisting of data recorded from all the three IoT products she owns, then it would be able to understand Jane (and also thousands of other similar users) better and optimize its product supply chain. Such optimization would allow *TastyCoffee* to reduce costs and wastage, which would increase profits. Further, such data would help *TastyCoffee* to improve its product lines and introduce new products to the market rapidly, which would also lead to a strengthening of its brand value. From Jane's perspective, the rewards (e.g. voucher, discounts) received from *TastyCoffee* would motivate her to participate and trade data in the S$^2$aaS model.

Jane primarily uses these IoT solutions due to the importance of their primary functions. She has a smart coffee machine that automatically brews coffee when she gets up in the morning so, by the time she arrives in the kitchen, coffee is ready for her. Moreover, She has a smart activity monitor, *Fitbit*, which monitors her exercise patterns, food intake, step counts, goals, and so on. Finally, Jane uses *Beddit* to monitor her sleep pattern and quality. These IoT solutions are manufactured by different companies, and they work independently. However, S$^2$aaS ecosystem would allow both Jane and external entities to trade data in exchange for rewards.

Data consumers may also use S$^2$aaS to fulfil very complex data requirements. We believe that to gather whatever data, or just *'Big Data'*, does not have much value in some circumstances. Let us consider the following data requirement:

- *collect ECG, SpO$_2$ (Oxygen level in blood) data, activity data, and sleep data where the data owners follow the daily pattern of sleeping at least 6 hours, drink coffee more than 2 times per day, go to gym in the evenings and at least go to gym 3 times a week, constrained to certain geographical location (and maybe age group restrictions as well).*

In today's IoT ecosystem, there is no way to fulfil data requirement such as above. However, with the help of different IoT solutions, S$^2$aaS could fulfil such queries. It is important to note that, in some circumstances, data captured by IoT solutions can be used to validate data collection conditions (see the above data request) and in other circumstances, data captured by IoT solutions become the result set to data requests.

In a similar line of thinking, let us consider a data request by a data consumer related to a smart city domain [10]:

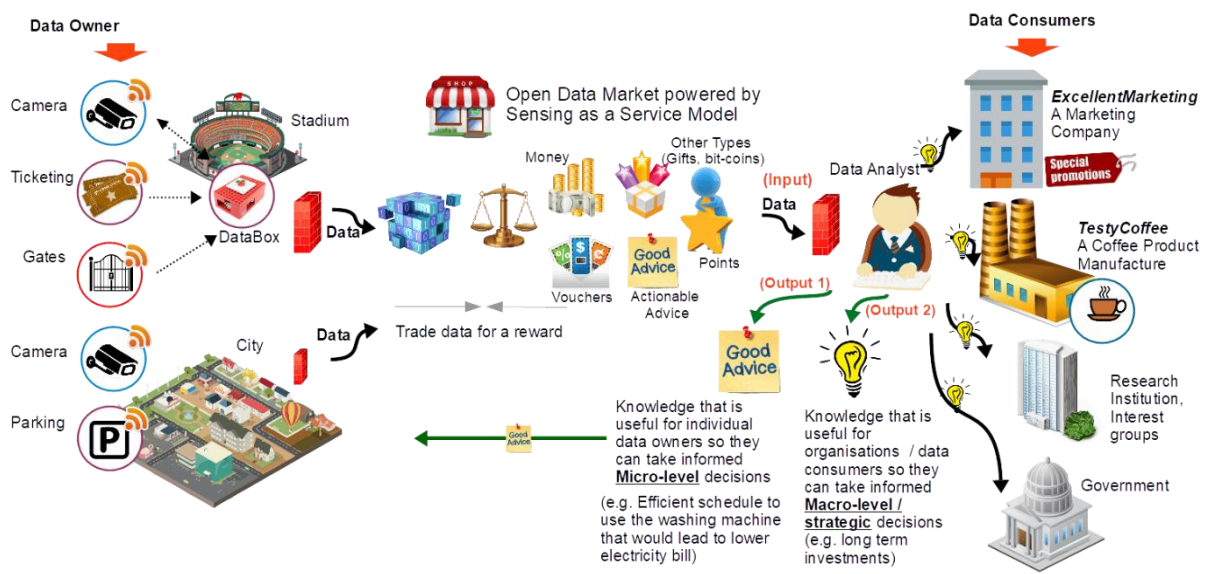

*Figure 2: Data Market for Sensing as a Service in Smart City Domain*

- *Collect crowed movement data, weather data, parking data, gender ratio and age groups of people who attend a local event in the 'Newcastle City' stadium when there is a different national event streamed/telecasted in online / TV.*

This data request is coming from a data consumer called *ExcellentMarketing*, a private firm specialized in viral and efficient promotional campaigns. In today's IoT ecosystem, there is no way to fulfil data requirement such as above. However, reward-based S$^2$aaS would encourage private and public entities to make their data available to external entities to be requested and traded in exchange for rewards. S$^2$aaS makes it very easy with small budgets to collect large volumes of, but more importantly most relevant data, not just big data.

It is important to note that, in this paper, we only focused on data market places where personal data is somehow involved. For example, in both use cases we discussed above, there is a single individual who owns the personal data. However, we would like to acknowledge that there is an entire segment of industrial data trading where either public or private entities own data, but not individuals. Some of the leading companies have already started creating industrial data market places allowing business to create new revenue streams using their existing sensing infrastructure[2]. In this paper, we consider such industrial data trading out of the scope.

**ARCHITECTURE AND COMPONENTS**

Let us now present the main components of the S$^2$aaS ecosystem. It is important to note that, based on reliability, security, and privacy expectations, the actual architecture may be varied. Today's mobile app ecosystem has largely inspired our design thinking. We identify five major components: 1) Data Bucket, 2) Data Market, 3) Data Studio, 4) Data Mill, and 5) Data Oven.

**Data Bucket:** This service is responsible for gathering data from different IoT solutions. It interfaces with the data owners using a mobile app where it allows data owners to express their privacy preferences, receive recommended data requests, trade data, and negotiate data trading [11]. We will walk you through the **Data Bucket app** later in this paper in order to demonstrate how data owners may engage with the S$^2$aaS ecosystem. Each data owner has its own data bucket.

**Data Market:** This service is similar to the Google Play app store. Instead of apps, Data Market stores organises, disseminates, and manages data requests. Data Market also organises and manages Metadata provided by individual Data Buckets. Such Metadata allows Data Market to distribute data requests appropriately to compatible and interested data owners.

**Data Studio:** This service allows data consumers to create their data requests easily. It provides necessary integrated development environment like interface that allows data consumers to compose data requests efficiently and effectively. Each data request is a package of several pieces of information that includes data requirements, rewards, privacy risks, analytical components, and other information. It is important to note that having a standardization of data format for data exchange is vital for the success of this module. There are some ongoing efforts in this area such as IoTivity (iotivity.org), Hypercat (hypercat.io), and AllJoyn (openconnectivity.org).

**Data Mill:** This is a technical infrastructure service component where personal data is being processed in combined with the open data (e.g., weather data). No stakeholders involved with this component directly. A brand-new milling machine is created in order to gather and process data from a single data owner. It is not allowed to combine personal data from different individuals within a single machine mill [12]. Data Mill either could be located in the cloud or within the local device within the smart home (e.g., as part of Amazon Echo or Google Home) [13]. We envision that, in the future, companies would prefer to process the data locally on the edge of the network, to avoid (or minimise) any legal challenges and hefty GDPR related penalties.

---

[2] https://networks.nokia.com/services/sensing-as-a-service

**Data Oven:** This is also part of the technical infrastructure service. It receives data from the Data Mill where initial data processing occurs. Data from multiple different individuals are processed together within the Data Oven.

Figure 3 illustrates the high-level architecture of the S$^2$aaS ecosystem including some of the most important communication aspects.

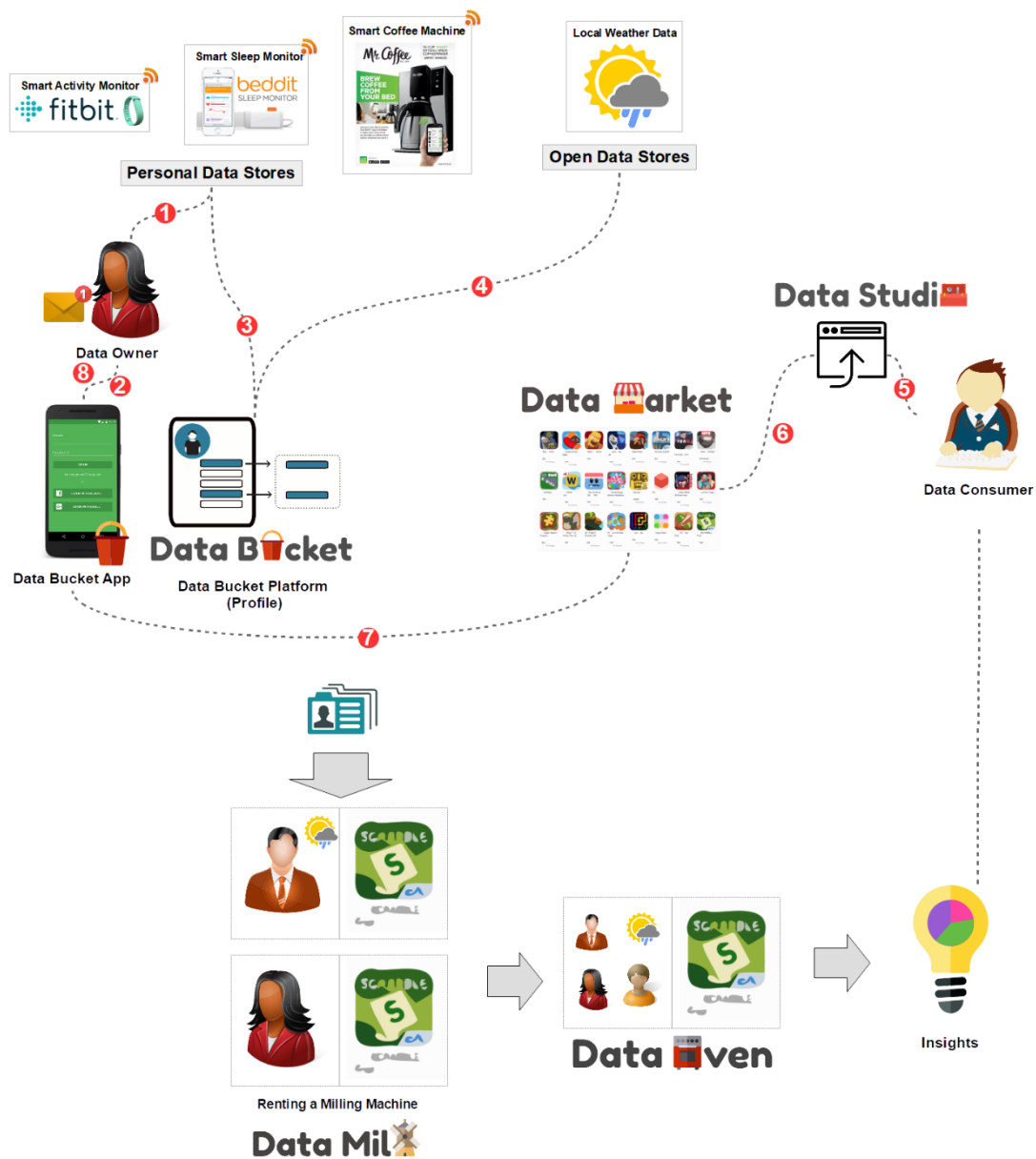

*Figure 3: Major components in the Sensing as a Service Ecosystem*

The Sensing as a Service model builds on top of the existing IoT ecosystem. The IoT solutions are typically registered and configured by data owners. These IoT solutions communicate with their companion cloud services, and data is frequently being pushed back in order to be analysed and knowledge extracted. Data owners need to register themselves by creating and configuring a data bucket account. Once logged in the data bucket app, data owners are provided with user interfaces that allows them to connect IoT solutions to the data Bucket.

In the example, Jane has connected *Fitbit*, *Beddit* and *Smart coffee machine* to her Data Bucket account. During each of these configuration processes, data owners are allowed to express their preferences in terms of which data items they would like to trade under what conditions, and so on.

Let us now look at the other end of the $S^2$aaS ecosystem, the data consumers. Data consumers first need to do some preliminary research and determine kinds of data they need to gather in order to support their objectives. In the sample scenario, the data consumer needs to research activity, sleep patterns, and their relationships with coffee consumption.

The data consumer then needs to use the Data Studio to create the data request. Data request comprises of several pieces of information including data items requested, the intention of data gathering, knowledge expected to be derived, technologies used to processed and analysed, rewards willing to provide, and so on. The Data Studio packages this request and publishes it in the Data Market place. The Market place then pushes it to the matching data owners.

Depending on the initial configuration, data owners will either receive the data request as a notification or will be listed under the recommended data trading section in the Data Bucket app. Data owners may open up the request to continue trading data. Data owners can use the Data Bucket app to negotiate with the respective data consumers regarding rewards, and the exact data to be traded (e.g., data granularity, duration, etc.). We present some examples later in the paper. Once both the data owner and the consumer are agreed, a digital contract will be made.
Data market passes the authorization to gather data (as per the agreement) to the Data Mill in order to perform data processing. At the same time, data owners receive the agreed reward. In the Data Mill, personal data is processed in combination with the data gathered from open data sources (e.g., public data such as weather). Once completed, processed data is sent to the data oven in order to perform further processing, analysis, and derive expected knowledge. At this stage, data from multiple users are processed together.

**ENVISIONING ECOSYSTEM AND CHALLENGES IN USER ENGAGEMENT**

Let us now go through the major user interfaces provided by the Data Bucket app in order to explain how a typical non-technical user may participate with the $S^2$aaS ecosystem. Our intention is not to make the UI designs perfect. Instead, we aim to envision the high-level objectives of each screen. These interfaces allow us to highlight challenges in the user interaction design.

4 (a) shows the login screen of the Data Bucket app. Data Bucket is the central account for data owners who interact with the $S^2$aaS ecosystem. Each Data Bucket is registered with one or more Data Markets. Once logged in, as shown in 4 (b), a list of IoT products are shown. Data owners can click the IoT solutions they own and configure them. As shown in 4 (c), Data Bucket app provides an interface for data owners to enter their credentials related to each IoT solution (e.g., login details for the Fitbit account).

Data Bucket knows the exact data items that a particular IoT solution can provide (e.g., Fitbit provides body weight, physical activity, step count, body mass index, and sleep duration). Data owners can then select which data items they would like to trade and several other preferences. Using a similar process, data owners can connect different IoT solutions. As a result, Data Bucket knows which data items are available for trade by each data owner.

In a separate screen 4 (d), data owners are provided with a list of recommended data trading offers. Data trading opportunities can be broadly categorised into two, namely, 1) one-time, and 2) subscriptions. Once data owners decide to explore further with any of the trading offers, they are provided with a secondary screen, as shown in 4 (e).

Data owners are provided with details of a particular data-trading offer (who is the data consumer, what is the intention, what analytics are used, how analytics are certified, what knowledge is expected to derive, and so on). Data owners can negotiate how much data they want to trade under which conditions and how much reward they would expect in return.

Once both parties agreed, details can be seen on different screen, as shown in 4 (f). This screen allows data owners to claim their rewards (see 4 (g)) and cancel existing subscriptions. One of the important types of reward is actionable (useful) advice. Instead of giving financial rewards, data consumers agree to provide useful advice to the data owners through a designated app or through the Data Bucket app's insights screen as shown in 4 (h).

Typically, data owners are non-technical personnel. Therefore, the above-mentioned user interfaces and interactions should be built in such a way that they can be used with minimum technical knowledge. The challenge is to evaluate data requests made by data consumers and generate risk-reward analysis reports so the data owners can make informed data trading decisions. Visually representing risk-reward analyses in such a way that they are detailed enough for data owners to be informed accurately, but simple enough to be understood easily and quickly is an important feature towards the success of the S²aaS model. One of the challenges is to determine what information is

| AN ANALOGY |
|---|
| *"You can think of the data consumer as a baker (joe the friendly baker). Let us assume he wants to make a new coffee cake. He first builds the recipe using his recipe book (**Data Studio**). Once he is happy with the recipe, he goes to the market (**Data Market**) and buys coffee beans, wheat, sugar, eggs, and all other ingredients (**Data**). Then, joe goes to his village mill (**Data Mill**) and rent three different milling machines that are designed for different grinding requirements, one for wheat, one for sugar, and one for coffee. Joe had to wait until few other village men finish their grindings and release the milling machines. Important rule in the mill is that each batch from each customer need to be grinded separately. However, depending on the grinding requirements, customers can pick a specialised machine. Once all done, joe takes all ingredients to his bakery. He combines all the ingredients according the recipe he built earlier. Joe bakes his cake in his oven (**Data Oven**) until he satisfies with the outcome. Coffee cake (**Insights**) is ready!"* 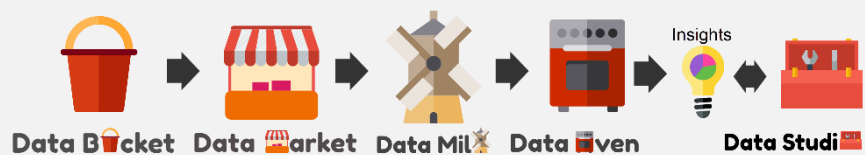 |

important for each data owner when engaging with data trading and how such information can be presented to them.

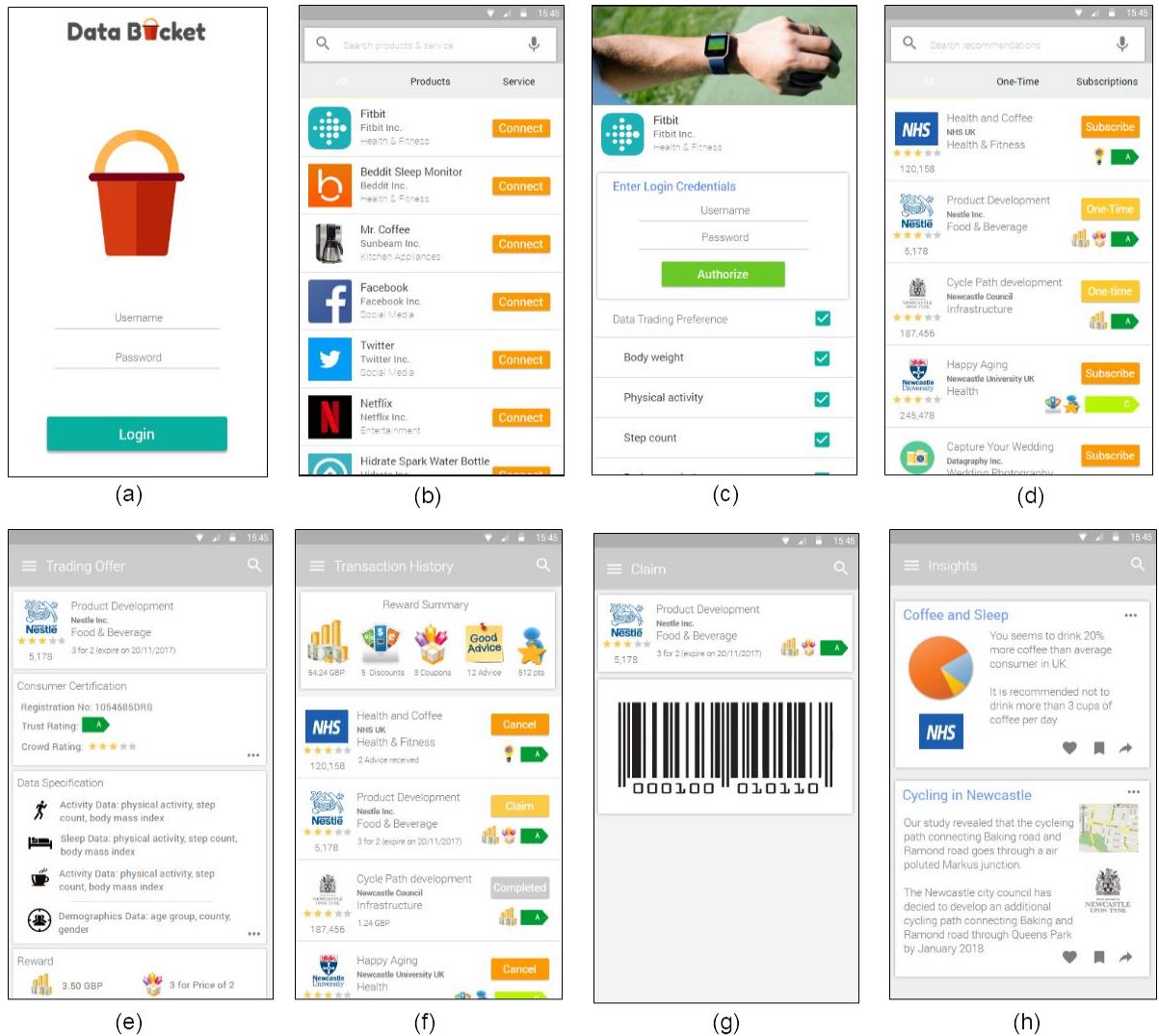

Figure 4: Data Bucket Envisioned User Experience

Another challenge is to decide what kind of controls should be given to data owners during both the negotiation and post-trading stages. The data buying and selling processes should be simple enough to take place repeatedly without requiring significant amounts of input and time from data owners. Finally, what aspects of a data trading transaction are negotiable and unnegotiable is also an important question. Baarslag et al. [14] have provided some insights towards data trading negotiations.

Before concluding this paper, it is worth to briefly look at existing IoT platforms and their position concerning the $S^2$aaS model. In summary, no platform currently supports personal data marketplaces (see Table 1).

Table 1: Current IoT platforms and their position concerning the S2aaS model.

| Platform | Position |
|---|---|
| ThingSpeak (thingspeak.com) | All these platforms are designed to help build IoT applications quickly. They typically provide mechanisms to communicate with IoT devices (e.g. API). Additionally, they would provide common features such as data storage, analytics, visualizations, events triggers, and notifications. These platforms are designed to assume that the data they manage is owned by the IoT application owners. There isn't any data marketplace concept built into the design. |
| Xively (xively.com) | |
| Udidots (ubidots.com) | |
| OpenIoT (openiot.eu) | This is the only platform that supports $S^2$aaS model by design. However, its primary focus is industrial data. More specifically, it addresses the problem of semantic data modelling where data can be queried over heterogeneity of data sources without knowing underline hardware and software structures in the context of IoT. Further, OpenIoT is focused on data acquisition for free. No pricing mechanisms are implemented into the platform. |

**CONCLUSION**

Throughout this paper, we discussed why sensing as a service has the capability to fix the weakness in existing IoT ecosystem. It also addresses both social and economic issues related to IoT by bringing the data ownership back to the owners and generating value for all the stakeholders. Today, we see a glimpse of data trading efforts. For example, Google Opinion Reward and Survey.com are mobile applications that selectively present survey questionnaires to users. Users are paid for answering the questionnaire surveys. Survey questionnaires have issues such as accuracy of answers, difficulty in asking lots of questions (users get bored quickly, despite the fact that they are getting paid), difficulty in getting answers due to the fact that users may not remember (e.g. how many times did the user drink coffee over the last month), and so on. Therefore, we can imagine how much value is hidden in the data captured by different IoT solutions. Sensing as a Service ecosystem allows us to liberate our data from large corporates where we do not have much control over the data or what they would potentially do with our data. Further, the $S^2$aaS ecosystem would bring down the data acquisition cost significantly. In conclusion, $S^2$aaS adds some level of discipline towards personal data collection and usage by third-party entities, while it brings both controllability, accountability, and fairness (e.g., rewards) to the IoT ecosystem.

**ACKNOWLEDGEMENTS**

This work is partly supported by EPSRC PETRAS 2 (EP/S035362/1).